\documentclass[preprint]{aa}
\usepackage{graphicx}

\usepackage[varg]{txfonts}
\usepackage{cases}
\usepackage{natbib}
\usepackage{underscore}
\usepackage{hyperref}

\usepackage{amsmath}

\begin{document}
      \title{Dependence of the intensity of the nonwave component of EUV waves on coronal magnetic field configuration}
      
      \author{Yuwei Li\inst{1}, J. H. Guo\inst{1}, Y. W. Ni\inst{1}, Z. Y. Zhang \inst{1}, P. F. Chen\inst{1,2}}
      
      \institute{Key Laboratory of Modern Astronomy and Astrophysics, School of Astronomy and Space Science, Nanjing University, Nanjing 210023, PR China \\
      	\email{chenpf@nju.edu.cn}
 	    \and 
 	    State Key Laboratory of Lunar and Planetary Sciences, Macau University of Science and Technology, Macau 999078, PR China
 	    }
    \titlerunning{Intensity anisotropy of EUV waves}
 \authorrunning{Li et al.}
\date{}

\abstract{Mounting evidence has shown that EUV waves consist of a fast-mode magnetohydrodynamic (MHD) wave (or shock wave) followed by a slower nonwave component, as predicted by the magnetic fieldline stretching model. However, not all observed events display both wavefronts, particularly the slower nonwave component. Even in case that the slower nonwave component is present, the intensity distribution often exhibits strong anisotropy.}
{This study is intended to unveil the formation condition of the slower nonwave component of EUV waves.} 
{We analyzed the EUV wave event on 8 March 2019, and compared the EUV wave intensity map with the extrapolation coronal potential magnetic field. Data-inspired MHD simulation was also performed.}
{Two types of EUV waves are identified, and the slower nonwave component exhibits strong anisotropy. By reconstructing 3D coronal magnetic fields, we found that the slower nonwave component of EUV waves is more pronounced in the regions where magnetic fields are backward-inclined, which is further reproduced by our MHD simulations. }
{The anisotropy of the slower nonwave component of EUV waves is strongly related to the magnetic configuration, with backward-inclined field lines favoring their appearance. The more the field lines are forward-inclined, the weaker such wavelike fronts are.}

\maketitle
\nolinenumbers

\section{Introduction}\label{introduction} 

Solar extreme-ultraviolet (EUV) waves are a spectacular and frequently observed large-scale wave phenomenon associated with other solar eruptions, such as solar flares, filaments eruptions, and coronal mass ejections \citep[CMEs,][]{Moses1997, Thompson1998, Chen2006, Nitta2013, mei20}. In the early time, since EUV waves often propagate in a quasi-isotropic manner \citep{Moses1997, Thompson1998}, and are occasionally co-spatial with H$\alpha$ Moreton waves \citep{Thompson2000, Pohjolainen2001}, they were naturally considered to be fast-mode magnetohydrodynamic (MHD) waves or shock waves \citep{Thompson1998, Thompson1999, Wang2000, Wu2001, OfmanandThompson2002}. However, later studies performed by \citet{DelanneeAulanier1999} and \citet{Delannee2000} discovered that EUV wave fronts sometimes stop at magnetic separatrices. In addition, the speeds of EUV waves are typically three times smaller than those of Moreton waves \citep{Klassen2000, Zhang2011}, and sometimes below 100 km s$^{-1}$ \citep{zhuk09, grech22}, which are even slower than the coronal sound speed. All these features posed a big challenge to the fast-mode MHD wave model to account for some coronal EUV waves \citep{will09}.

In order to resolve the discrepancies, \citet{Chen2002} performed MHD numerical simulations, and found that as a flux rope erupts, there appear two EUV waves, i.e., a faster one and a slower one, the latter of which is immediately followed by expanding dimmings. The faster one is a fast-mode MHD shock wave piston-driven by the erupting flux rope, and the slower one is about three times slower than the faster one. To account for the formation of the slower-component EUV wave, they proposed the magnetic fieldline stretching model, i.e., it is an apparent propagation (or called nonwave) formed by the successive stretching of magnetic field lines pushed by the erupting flux rope. With a concentric semicircular magnetic configuration, they derived an analytical solution for the slow-component EUV wave, which is exactly about three times slower than the fast-mode wave speed. Such a feature is consistent with the statistical results \citep{Klassen2000, Zhang2011}. The prediction of two different types of EUV waves by \citet{Chen2002} were soon confirmed by \citet{Harra2003} with the Transition Region and Coronal Explorer (TRACE) observations. After the high-cadence observations of the Solar Dynamics Observatory (SDO) were available, more and more research supported the two-wave paradigm \citep{ChenWu2011, Schrijver2011, Asai2012, cheng12, Kumar2013, chan24, hu24}. Actually, even before the SDO era, there was already kinematical evidence for different classes of EUV waves \citep{warm11}, and only one class can be considered as fast-mode MHD waves \citep{temm11, vero11, pats12, kwon13, selw13, mann23}. It is now widely recognized that the EUV waves are composed of two components, including a fast MHD wave and followed by a slower nonwave component, with the slower wavefront historically referred to as the ``EIT waves" \citep{Chen2016}. Although the magnetic fieldline stretching model is widely accepted, other interpretations for the slower nonwave component also exist, including the current shell model \citep{Dalanne2008}, successive magnetic reconnection \citep{Attrill2007}, and the echo of the fast-mode shock \citep{Wang2009, Xie2019}.

Albeit this, not all solar eruptions manifest the simultaneous existence of both components, i.e., the leading fast-mode MHD wave (or shock wave) and the subsequent nonwave. For example, in the Solar and Heliospheric Observatory (SOHO) era, only the slower nonwave component can be detected in most events, which is due to the low cadence of its EUV telescope. According to \citet{ChenWu2011}, only when the observational cadence is less than $\sim$70 s, can the fast component be detected. However, even when the cadence is as short as 12 s in the SDO data, a single wave is discernible in some events. For example, in the event studied by \citet{Zheng2020}, only a single wave was observed to propagate with a speed of 500 km s$^{-1}$, which should correspond to the fast-component EUV wave according to the criteria proposed in \citet{Chen2016}. In contrast, the slower events in \citet{Nitta2013}, those with speeds below 300 km s$^{-1}$, probably represent the nonwave component of EUV waves only since the fast component is too faint. Moreover, even when the two components are visible in a single event, each wave generally exhibits strong intensity anisotropy. For the anisotropy of fast-component EUV wave fronts, it has been demonstrated that they are generally brighter near areas with weaker magnetic field. There are dual effects for it. First, a filament, once triggered to erupt, tends to propagate toward the direction with weaker magnetic field, and such inclined eruption further makes the piston-driven shock wave stronger on the side with weaker magnetic field \citep{Zheng2023}. Second, fast-mode MHD waves or shock waves tend to be refracted toward the area with weaker magnetic field \citep{uchi68, liu18, zhou24}. It is still unclear what is responsible for the anisotropy of the slow-component EUV waves.

In this paper, we analyze an EUV wave event driven by a filament eruption on 8 March 2019, wherein the intensity of the slow-component EUV wave exhibits strong anisotropy. By performing coronal magnetic field extrapolation and MHD simulations, we investigate the relationship between the intensity distribution of EUV waves and the local coronal magnetic field configuration, with the purpose to reveal the formation condition of the slow-component EUV waves. The overview of observations is described in Section~\ref{sec:obs}, the coronal magnetic field reconstruction is shown in Section~\ref{sec:ext}, the simulation validation is presented in Section~\ref{sec:mhd}, followed by a summary and discussions in Section~\ref{sec:dis}.

\section{Observations and Data Analysis}\label{sec:obs}

\subsection{Instruments and Event Overview}

The event under study is a filament eruption associated with a C1.3-class flare on 8 March 2019 in NOAA active region (AR) 12734. Although the flare was very weak, this filament eruption excited distinct bright wave fronts. The filament eruption produced a CME and was accompanied by a type II radio burst, which is a typical CME-driven EUV wave event. This event was well documented by the Atmospheric Imaging Assembly (AIA) on board the SDO satellite \citep{Lemen2012}. The AIA instrument has 7 EUV channels and provides solar full-disk images with a spatial resolution of 0.6$^{\prime\prime}$ per pixel and a temporal cadence of 12 s. Typically, due to significant density compression at the EUV wave front, the wave front is clearer in the 193 \AA\ difference images. Therefore, in this study, we primarily selected the 193 Å channel for observation. To better track the propagation of the EUV wave, we extracted slices along different directions on the solar surface and considered the spherical projection effect, starting from the eruption source site. Furthermore, to reduce uncertainties in the observations, we set a certain width for each slice in each direction, which allows for a more accurate tracing of the EUV wave front motion \citep{Shen2012}.

\subsection{Observations of EUV waves}
\begin{figure*}
  \includegraphics[width=15cm,clip]{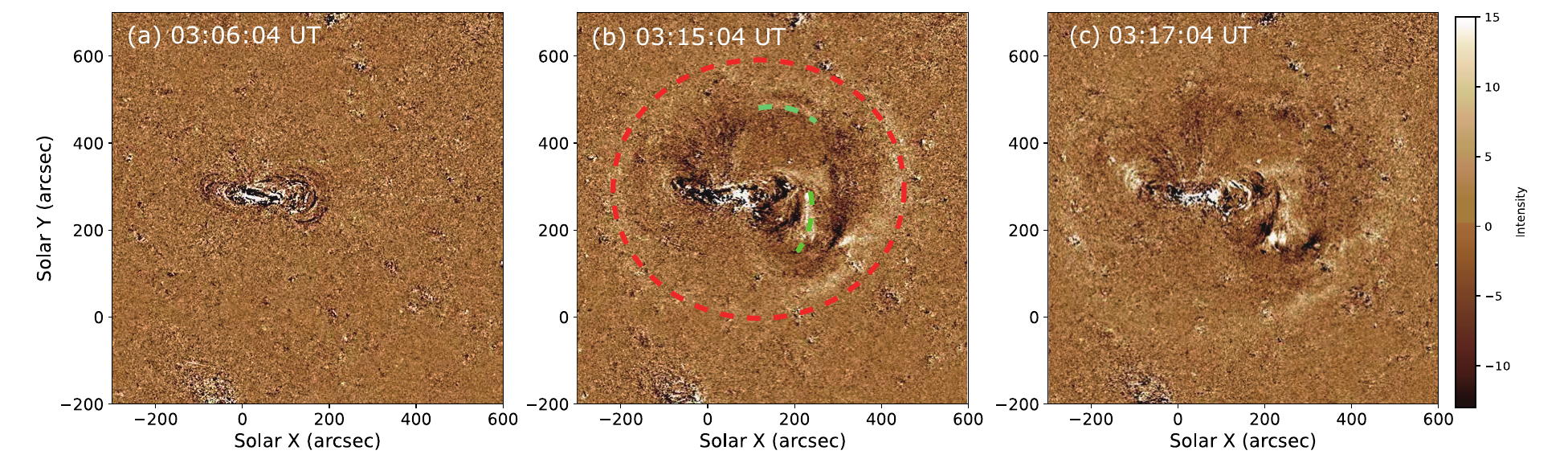}
  \centering
  \caption{SDO/AIA 193 \AA\ running-difference images illustrate the evolution of the EUV wave on 8 March 2019. The base time is 24 s earlier for each panel. The red and green dashed lines represent the two-component wavefronts in panel (b). The animation of this figure is available online. \label{figure1}} 
\end{figure*}

Figure~\ref{figure1} displays running-difference images in the 193 \AA\ channel, in which each image is processed with the Gaussian smoothing technique with a full width at half maximum being 3 pixels to reduce noises. This eruption produces two wavelike fronts, including a leading weak front (the red line in Figure~\ref{figure1}b) and an ensuing wave with patchy fronts (the green lines in Figure~\ref{figure1}b) followed by dimmings. The leading wave, with a roughly elliptic shape, is rather isotropic as indicated by the red dashed line in Figure~\ref{figure1}b. The ensuing wave is more anisotropic, being extremely bright in the southwest direction, and marginally visible in the north direction, but nearly invisible in the south direction, as indicated by the green dashed line in Figure~\ref{figure1}b.

\begin{figure*}
  \includegraphics[width=15cm,clip]{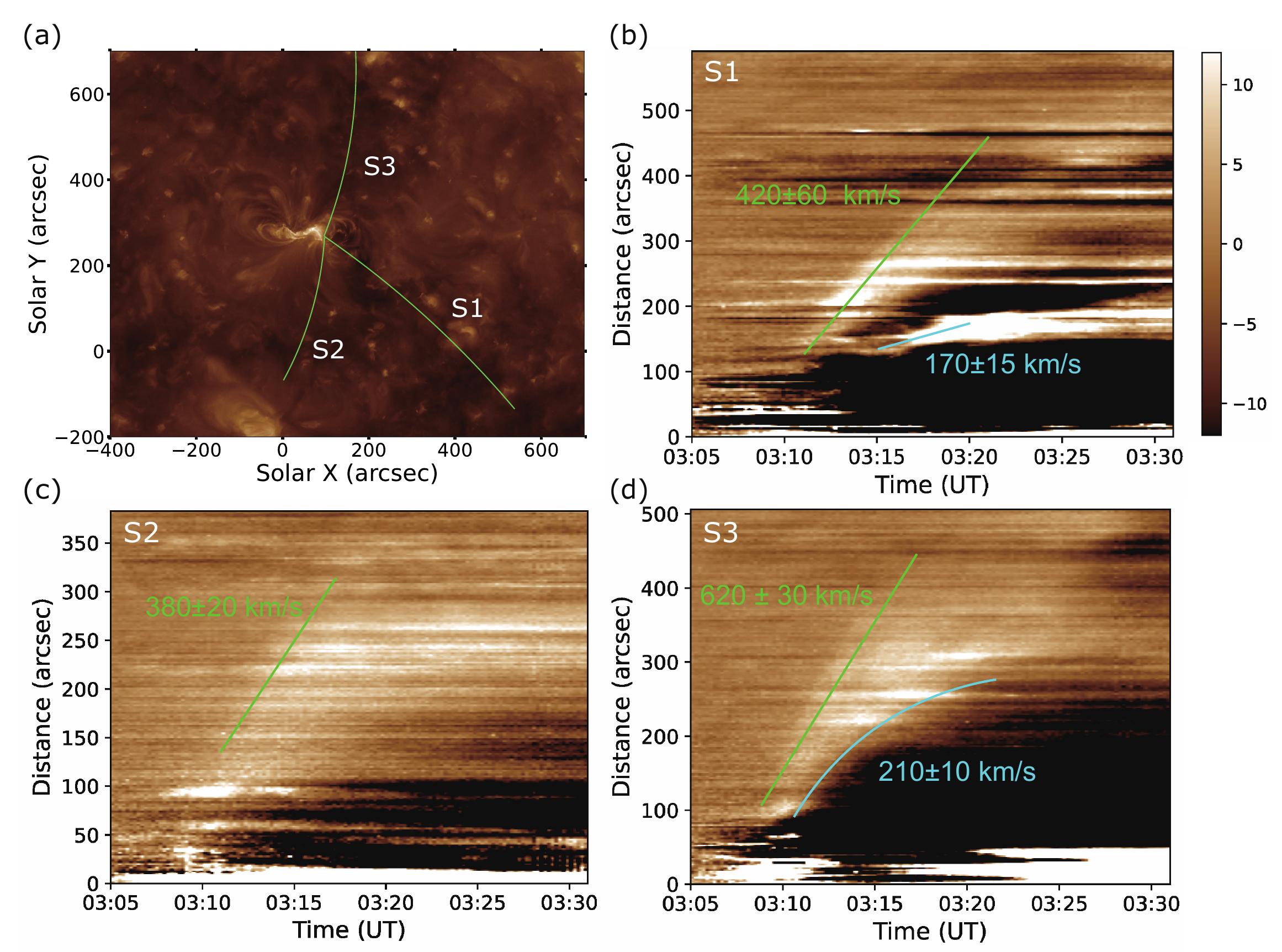}
  \centering
  \caption{Time-distance diagrams of the base-difference intensity at EUV 193 \AA\ channel along slices S1 (southwest, panel b), S2 (south, panel c), and S3 (north, panel d). The slice trajectories are marked in panel (a). Two wavefronts and the fitting speeds are shown in panels (b) and (d), while only one wave is visible in panel (c). Note that both the fast MHD wave component (green lines) and the slower nonwave component (cyan lines), accompanied by pronounced expanding dimmings, are visible along slices S1 and S3, whereas only the fast MHD wave component can be clearly identified along slice S2. The error bars are estimated from the upper and lower boundaries of wavefronts. \label{figure2}}
\end{figure*}

To quantitatively investigate the kinematics of the EUV waves, we select three slices (the green lines in Figure~\ref{figure2}a) to generate time-distance diagrams, as shown in Figures~\ref{figure2}b--\ref{figure2}d. The green and blue oblique lines outline the wavefronts propagating along each slice. In Figure~\ref{figure2}b, the first wavefront appears around 03:10 UT along the southwest direction of slice S1, with a speed of $420\pm 60$ km s$^{-1}$. Subsequently, a second wavefront is observed around 03:16 UT followed by dimmings. It propagates at a slower speed of $170\pm 15$ km s$^{-1}$ initially, and soon it stops, appearing as a patchy EUV wave as investigated by \citet{Guo2015}. Figure~\ref{figure2}c shows the results of slice S2 in the south direction, where a wave can be identified to propagate with a speed of $380\pm 20$ km s$^{-1}$. Besides, at the distance of 250$^{\prime\prime}$, a stationary front is revealed, as discovered by \citet{chan16}. Such a feature was explained to be due to mode conversion from fast-mode wave to slow-mode wave, which is trapped at magnetic separatrix \citep{Chen2016, Chandra2018, Kumar2024}. Along the northward slice S3, Figure~\ref{figure2}d reveals two wave patterns. The first wave propagates out rapidly with a speed of $620\pm 30$ km s$^{-1}$, starting from around 03:09 UT. The other wave, which immediately borders the extending dimmings, propagates more slowly, and its speed gradually decelerates, with an average speed of $210\pm 10$ km s$^{-1}$. Comparing Figures~\ref{figure2}b--d, it is seen that the fast-component EUV wave is marginally isotropic, being slightly the strongest in the southwest direction, but the slow-component EUV wave is much more anisotropic, being most pronounced in the southwest direction and the weakest in the south direction.

\section{Relationship between 3D Coronal Magnetic Fields and EUV waves}\label{sec:ext}

It is expected that the properties of EUV waves are highly dependent on the coronal magnetic fields. Therefore, to explain the anisotropy of EUV waves in the spatial distribution, we reconstruct 3D coronal magnetic fields with the potential field extrapolation in the spherical coordinate, achieved by the method in PDFI$\_$SS library \citep{Fisher2020}. The computation domain extends from the solar surface to 2 solar radii, where the source surface is located.

\begin{figure*}
  \includegraphics[width=15cm,clip]{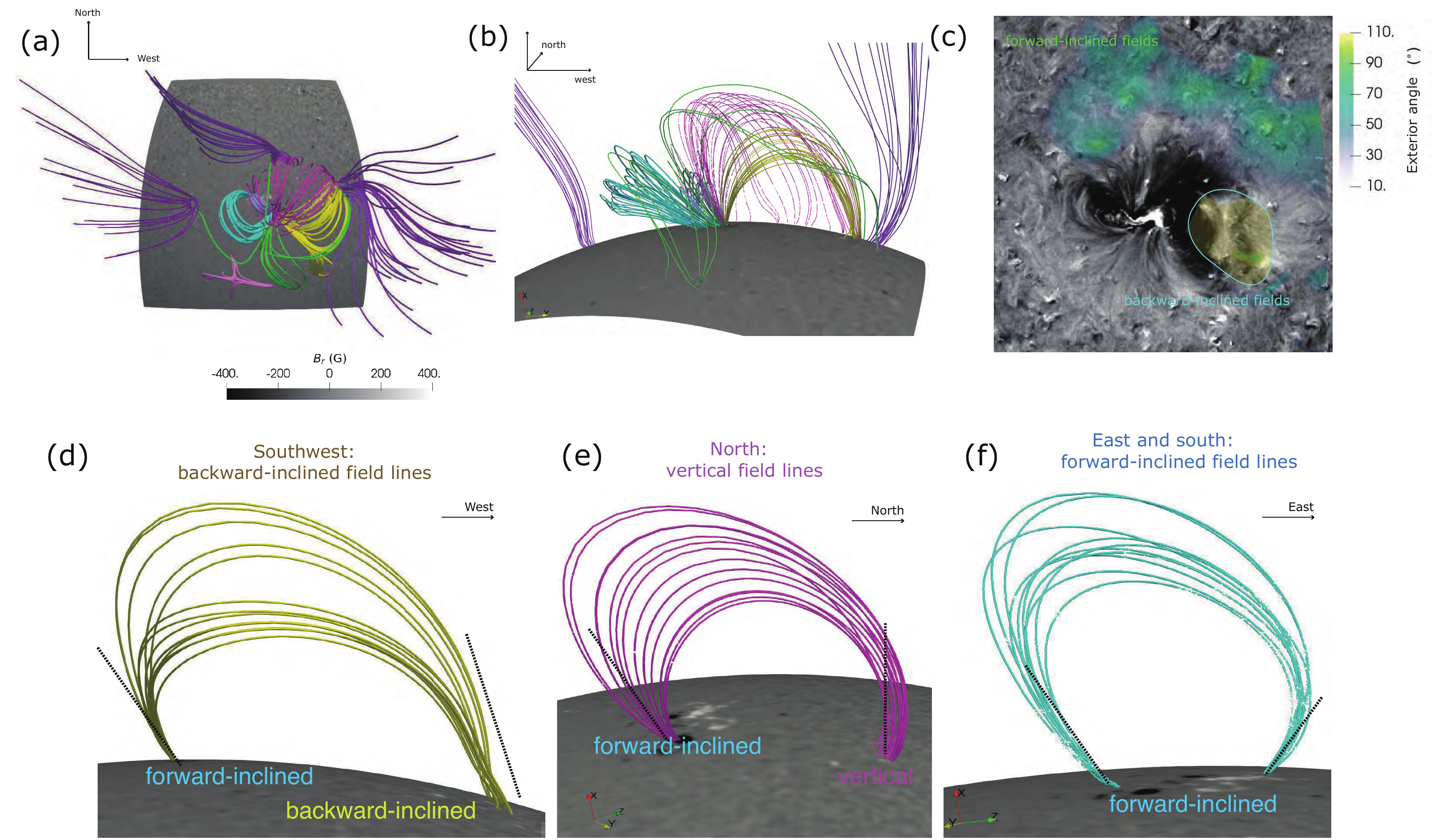}
  \centering
  \caption{Panels (a)--(b): Top view and side view of the 3D coronal magnetic field reconstructed with the potential-field extrapolation; Panel (c) displays the distribution of the exterior angle (color scale) overlaid on the SDO/AIA 193 \AA\ map at 03:15:04 UT; Panels (d)--(f) display some selected field lines, showing that those field lines traced from the western (yellow) and northern (magenta) regions are more backward inclined compared to the eastern (cyan) and southern (green) regions. \label{figure3}}
\end{figure*}

Figures~\ref{figure3}a--b show the 3D coronal magnetic-field lines from two perspectives. To investigate the relationship between the intensity of the slow-component EUV wave and the magnetic configuration, we define the exterior angle $\theta$ between a magnetic field line and the solar surface at the footpoint as follows:

\begin{equation}
	 \theta=\begin{cases}
	 	\arctan(B_r/B_t), & \text{if } B_r/B_t > 0; \\
	 	180^\circ + \arctan(B_r/B_t), & \text{if } B_r/B_t < 0,
	 \end{cases}
\end{equation}
\noindent
where $B_r$ is the radial component of the vector magnetogram and $B_t$ is the transverse component. With this definition, the magnetic field line is forward inclined when $\theta<90^\circ$, or backward inclined when $\theta>90^\circ$. We compute the exterior angle distribution of coronal magnetic fields with respect to the solar-surface plane in Figure~\ref{figure3}c, which is overlaid on the SDO/AIA 193 \AA\ image at 03:15:04 UT. It is seen that the exterior angle of the magnetic field lines, $\theta$, is larger than $90^\circ$ near the strongest patch of the slow-component EUV wave in the southwest direction. It is smaller than but close to $90^\circ$ in the north direction where the slow-component EUV wave is moderate. In contrast, $\theta$ is significantly smaller in the south direction, where the slow-component EUV wave is nearly invisible. It means that the intensity of the slow-component EUV wave is larger and larger when the exterior angle of the field line at the footpoint increases.

Based on the above analysis along with the 3D coronal magnetic-field extrapolation, it is found that the intensity of the slow-component EUV waves are strongly related to coronal magnetic fields, and is brighter at the backward-inclined magnetic field lines (e.g., the southwest direction in Figure~\ref{figure3}c). In the case of forward-inclined magnetic configuration, the more inclined the field line is, the weaker the slow-component EUV wave is (e.g., the south and north directions in Figure~\ref{figure3}c). Some selected magnetic field lines in the southwest, north, and south directions are displayed in Figures \ref{figure3}d--f.

\section{Data-inspired MHD Simulation}\label{sec:mhd}

\subsection{Numerical setup}
To validate the above conclusion extracted from observations and magnetic-field modelings, we conduct a 3D data-inspired MHD simulation to investigate how different configurations of background magnetic fields affect the intensity of slow-component EUV waves. To this end, we construct the initial magnetic-field configuration with different exterior angles of the background magnetic field lines on the two sides of a magnetic flux rope. Based on Titov-Demoulin-modified model \citep{Titov2014} (but slightly different from the original case with two magnetic charges), the initial background magnetic fields are provided by four sub-photosphere magnetic charges ($q_{1}=10\times10^{20}$ G cm$^2$, $q_{2}=-10\times 10^{20}$ G cm$^2$, $q_{3}= 1.3\times 10^{20}$ G cm$^2$, $q_{4}=-1.3\times10^{20}$ G cm$^2$), placed at the positions of (10, 0, -10), (-10, 0, -10), (120, 0, -100), (-10, 0, -100) Mm, respectively. The former two charges represent the fields of the core active region, while the latter two charges are adopted to imitate the large-scale magnetic fields. Then, a stable and force-free toroidal-shaped flux rope is inserted with the Regularized Biot-Savart Laws \citep{Titov2018} with a peak height of 10 Mm. In particular, to ensure the initial magnetic-field system being stable, we use the magneto-frictional method \citep{Guo2016} to relax it to a force-free state. The relaxed magnetic-field configuration is shown in Figure~\ref{figure4}a, from which one can clearly see that the field lines on the right side are more vertical (or even backward-inclined) than the left part, which is forward-inclined. That is to say, the exterior angle of the field lines on the right side is significantly larger than that on the left side.

The governing equations of our simulation are as follow:

\begin{align}
    \frac{\partial\rho}{\partial t} + \nabla \cdot (\rho \boldsymbol{v}) &= 0 \\
    \frac{\partial(\rho\boldsymbol{v})}{\partial t} + 
    \nabla \cdot (\rho\boldsymbol{v}\boldsymbol{v} + p\boldsymbol{I} - \frac{\boldsymbol{BB}}{\mu_0}) 
    &= \rho \boldsymbol{g}\\
    \frac{\partial \boldsymbol{B}}{\partial t} + \nabla \cdot(\boldsymbol{vB-Bv})= -\nabla \times(\eta \boldsymbol{j}) \\
    \frac{\partial e_{int}}{\partial t} + \nabla \cdot (\boldsymbol{v}e_{int}) &= - p \nabla \cdot
    \boldsymbol{v}+\eta J^{2}
\end{align}
where $\rho$, $\boldsymbol{v}$, $\boldsymbol{B}$, $p$, $e_{int} = p/(\gamma - 1)$, and $\boldsymbol{g}=-g_{0}e_{z}$, represent the density, velocity, magnetic fields, thermal pressure, internal energy, and gravity, respectively. In particular, we also adopt a uniform resistivity $\eta=2.0\times10^{-4}$ (normalized unit) for joule heating. Regarding the initial density and pressure, we adopt an isothermal atmosphere with a temperature of 1 MK, and then the density can be computed from the hydrostatic condition. The simulation includes two stages. First, to realize the energy balance, we relax the initial atmosphere coupled with magnetic fields until the density almost keeps constant. Then, to trigger the solar eruption, following \citet{Liu2024}, we implement a converging flow (the maximum velocity is 15 km s$^{-1}$) on  magnetic charges $q_{1}$ and $q_{2}$, thereby inducing magnetic reconnection below the flux rope and leading to eruption. This simulation is established on the MPI-AMRVAC framework \citep{Xia2018, Keppens2023}.

\subsection{Simulation results}
\begin{figure*}
  \includegraphics[width=15cm,clip]{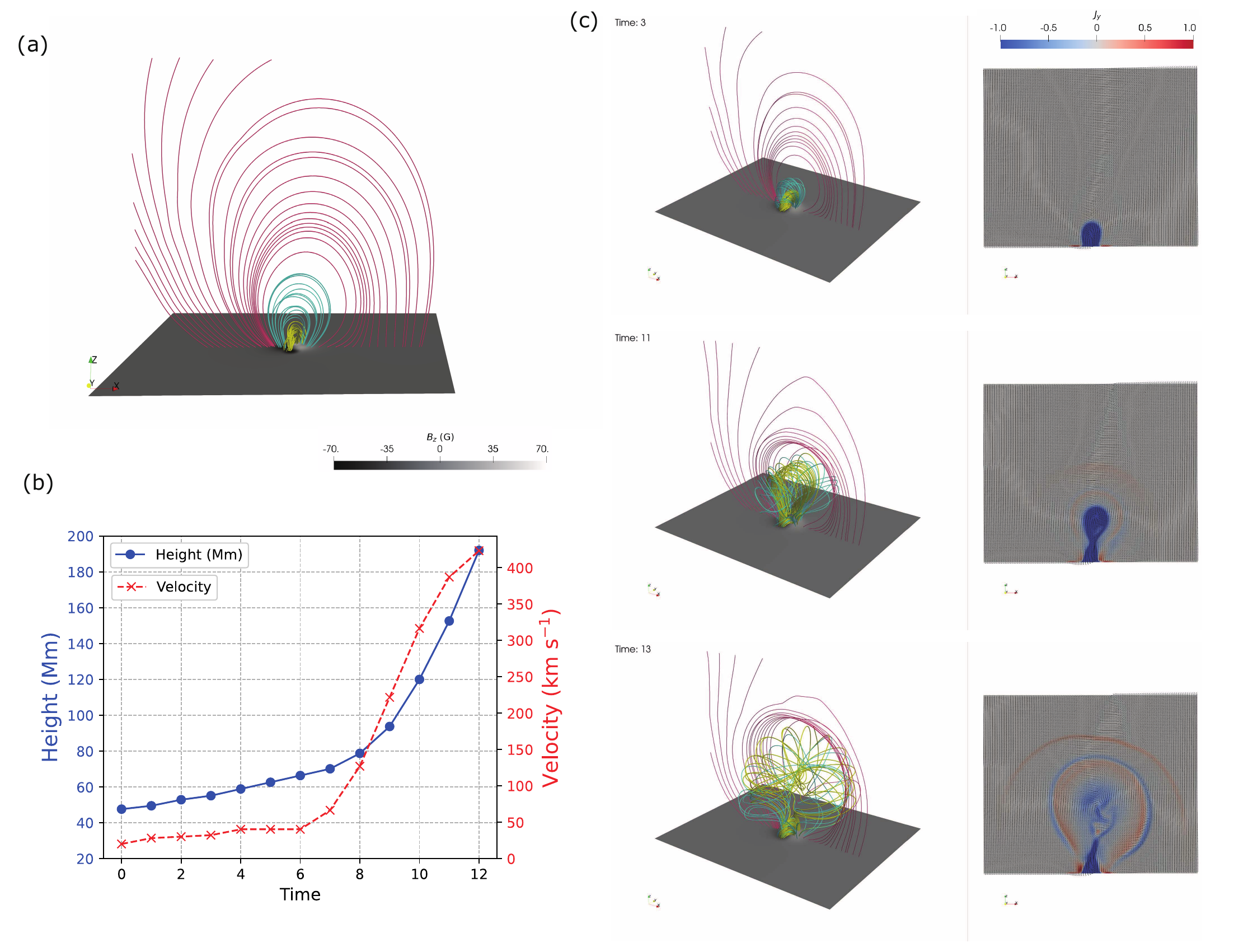}
  \centering
  \caption{Kinematics and 3D magnetic-field evolution of the eruption. Panel (a) shows the magnetic field configuration at the initial state. The yellow, cyan, and red-wine lines represent the preeruptive flux-rope filed lines, overlying active-region fields that are mainly generated by $q_{1}$ and $q_{2}$, and the large-scale background field lines extending from $q_{3}$ and $q_{4}$, respectively. Panel (b) shows the kinetics of the solar eruption, wherein the blue and red lines represent the evolution of flux-rope axis height and velocity, respectively. Panel (c) shows the evolution of 3D coronal magnetic fields and $J_{y}$ distribution in the $x$--$z$ plane at $t=3\tau_0$, 11$\tau_0$, and 13$\tau_0$. The animation of this figure is available online. \label{figure4}}
\end{figure*}

Figure~\ref{figure4} exhibits the kinematics (panel b) of the flux rope and the evolution of 3D magnetic fields and electric current (panel c) in the eruption process. As shown in Figure~\ref{figure4}b, the flux rope starts to rise with a velocity about 30 km s$^{-1}$ when the converging flow is imposed, which induces a quasi-static phase of the flux rope. When the flux rope slowly ascends to a height of 80 Mm at $t=6\tau_0$ (where $\tau_0=85.87$ s is the normalization time), the velocity rises up to about 130 km s$^{-1}$, during which a hyperbolic flux tube is formed and it triggers the slow rise phase of the flux rope \citep{Xing2024}. Hereafter, when the current-sheet becomes thin enough, fast reconnection is induced, which triggers the starting of the impusive phase \citep{Jiang2021}, after which a CME is formed and certain overlying magnetic fields (cyan lines) become part of the CME flux rope.

\begin{figure*}
  \includegraphics[width=15cm,clip]{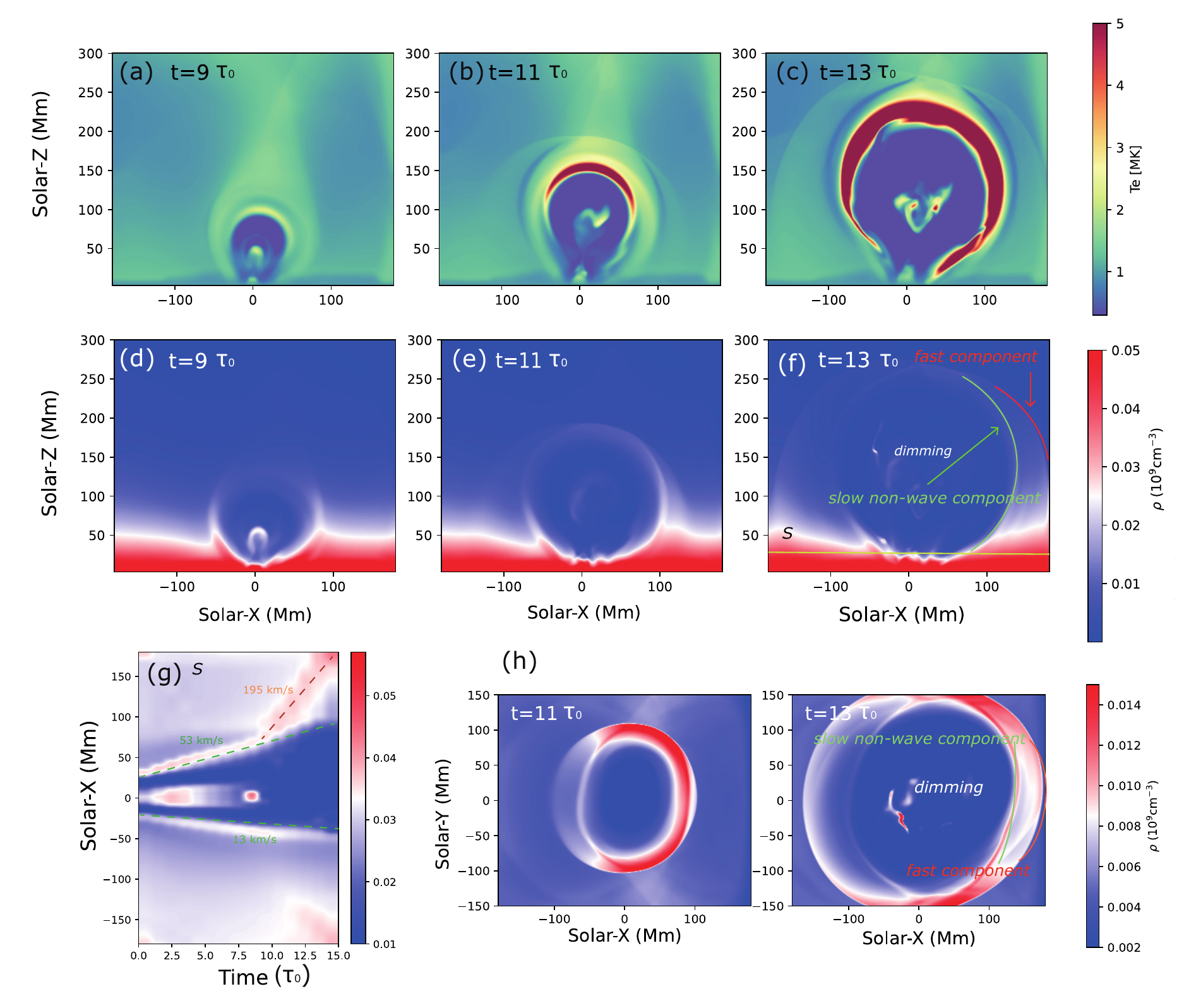}
  \centering
  \caption{Evolution of the plasma temperature and density during the eruption. Panels (a)--(c) show the distribution of the temperature in the $x$--$z$ plane at $t=9\tau_0$, 11$\tau_0$ and 13$\tau_0$, respectively. Panels (d)--(f) exhibit the evolution of the density in the $x$--$z$ plane at $t=9\tau_0$, 11$\tau_0$ and 13$\tau_0$, respectively. Panel (g) illustrates the time-distance diagram of the density along the slice $S$ in panel (f). Panel (h) shows the distribution of the density in the $x$--$y$ plane at the height of 150 Mm. The animation of this figure is available online. \label{figure5}}
\end{figure*}

Figure~\ref{figure5} shows the evolution of the plasma temperature and density in the eruption process. Figures~\ref{figure5}a--c illustrate the temperature distribution in the $x$--$z$ plane at $t=9\tau_0$, 11$\tau_0$, and 13$\tau_0$, respectively. A dome-like area is heated to 10 MK due to Joule heating around the cavity, corresponding to the boundary of the flux rope. In front of it, a moderately heated region can be recognized ($\sim$ 2 MK), which is due to the plasma compression and is relevant to the stretching of field lines \citep{Chen2002}. Figures~\ref{figure5}d--f show the density evolution, from which one can clearly identify a low-density cavity in the middle and two wavelike density-enhanced fronts further out, corresponding to the dark cavity, CME leading front and the shock wave in CME observations \citep{Chen2009, Guo2023}. To investigate the kinematics of the wavefronts, we cut a slice in the $x$-direction at the height of 30 Mm (slice S in Figure~\ref{figure5}f), and the time-distance diagram is displayed in Figure~\ref{figure5}g. One can see two wavefronts propagating in both directions. Along the right direction, i.e., $x>0$, a slow-component wave, which is indicated by the green dashed line, propagates with a speed of 53 km s$^{-1}$. Such a wave pattern is immediately followed by expanding dimmings. At $t=\sim 8\tau_0$, the impulsive acceleration of the flux rope drives a fast-mode shock wave, propagating outward with a speed of 195 km s$^{-1}$, as indicated by the red dashed line. The speed of the slower nonwave component is $\sim$3.68 times smaller than the fast-component wave. These results are consistent very well with the prediction of field-line stretching model \citep{Chen2002}, though the magnetic field is assumed to be weak for the simplicity of numerical simulation so that the speeds of both waves are near the lower limits in observations. Toward the left direction $x<0$, we can also identify two waves, but both are significantly weaker than in the right direction.

The time-distance diagram reveals that the wavefronts in the simulation exhibit a strong anisotropy due to the difference of overlying magnetic fields. To make it clearer, we display the density distribution in the $x$--$y$ plane at $z=150$ Mm, it can be seen that both wavefronts are stronger on the right side compared to the counterparts on the left side. As such, our simulation results are consistent very well with the findings from observations and coronal magnetic-field extrapolations.

\section{Discussions and Conclusion} \label{sec:dis}

Waves in the solar corona not only carry energy \citep{russ13}, but also serve as probes to diagnose coronal information that cannot be measured directly, e.g., magnetic field and temperature \citep{Ballai2007, West2011, Long2013, pian18, down21, naka24}. In this sense, EUV waves are particularly useful since they are generally observed to consist of a leading fast-mode MHD wave and a following slower nonwave component, with the latter commonly named as ``EIT wave". On the one hand, compared to local standing waves trapped in coronal loops or propagating waves in open flux tubes, large-scale EUV waves occupy the indispensable statue in diagnosing global coronal magnetic fields. On the other hand, while the fast-component EUV waves tell us the strength of coronal magnetic fields, the slow-component EUV waves reveal the coronal magnetic topology, including magnetic geometry and separatrixes \citep{Chen2009, Chen2016}. As such, a comprehensive understanding of the nature and formation condition of EUV waves is the premise for their applications in coronal seismology.

\begin{figure*}
	\includegraphics[width=12cm]{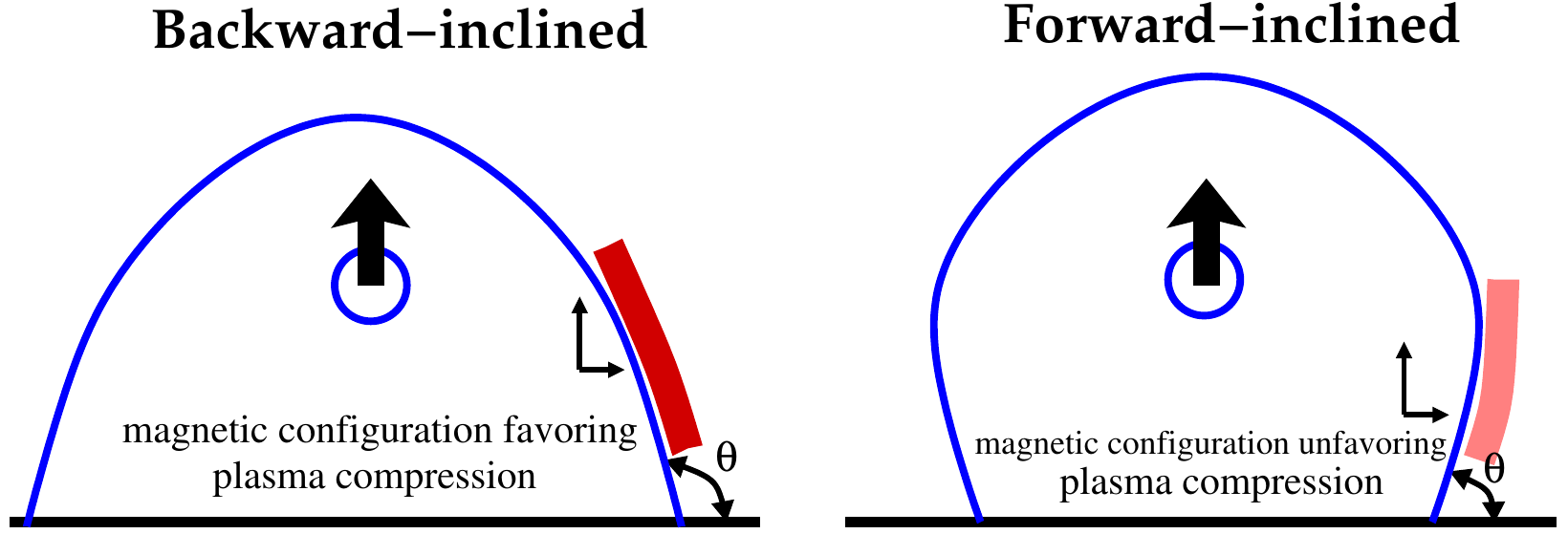}
	\centering
	\caption{Schematic sketch explaining why the backward-inclined magnetic field lines results strong plasma compression (left panel) and the forward-inclined magnetic field lines result in weak plasma compression (right panel), where the blue lines are magnetic field lines, $\theta$ is the exterior angle of the field lines, the dark red front means a brighter nonwave component front, and the light red front means a faint nonwave component front. \label{figure6}}
\end{figure*}

The two-wave paradigm of EUV waves is consistent with the theoretical prediction of the magnetic field-line stretching model \citep{Chen2002}, which states that, during the eruption of a magnetic flux rope, a fast-mode piston-driven shock wave propagates as the outermost front. Behind this shock wave, a slowly-propagating disturbance is generated by the ongoing stretching of magnetic field lines pushed by the flux rope. For each magnetic field line, the stretching propagates from the top part to the footpoints, compressing the plasma on the outer side of the field line to form emission enhancement, and the successive stretching of the filed lines creates an apparent wave phenomenon, i.e., ``EIT waves". It is noted that such stretching would naturally generate a current layer \citep{wong19} and sub-Alfv\'enic propagation of magnetic reconfiguration \citep{pasc17}. Under the assumption of a concentric semicircular magnetic field configuration, the speed of the slow-component wave is about one third of the leading fast-mode wave. However, frequently the wave phenomena in real observations are more complicated than the canonical paradigm, presumably due to inhomogeneity of coronal magnetic fields and plasma. For example, not all events exhibit both wave components, i.e., some events have the fast component only \citep[e.g.,][]{kouk20, hou23}, whereas others have the nonwave component only. In addition, even when both components are detectable, their intensity may show anisotropy. Hence, it is demanding to understand what are responsible for the anisotropy of EUV waves, and as an extreme case, why one component is missing in some events or along some directions. For this purpose, we investigated the EUV waves associated with a flare/CME on 8 March 2019. In this eruptive event, we observed both components of EUV waves, with the slow-component EUV wave displaying pronounced anisotropy.

The slow-component wave, or ``EIT wave", is extremely bright in the southwest direction to the source active region, moderate in the north direction, and much weaker in other directions, as illustrated by Figure \ref{figure1}. We examined the extrapolated coronal magnetic fields and compared the difference of the magnetic field at various locations. It is found that the magnetic field lines around the bright ``EIT wave" are more backward-inclined as indicated by Figures \ref{figure3}b--c, whereas the field lines in the north and southern directions are forward inclined, as seen from Figures \ref{figure3}e--f. Such a correlation is expected from the magnetic field-line stretching model \citep{Chen2002}. According to this model, the slow-component EUV wave is generated as the closed field lines overlying the erupting flux rope are pushed to stretch up. The stretching of the closed field lines is associated with two motions: upward motion as the eruption and lateral motion as the expansion. When the leg of the closed field line is more backward-inclined, as illustrated by the left panel of Figure \ref{figure6}, both motions would compress the outer plasma straightforwardly. On the other hand, when the leg of the closed field line is more forward-inclined, as illustrated by the right panel of Figure \ref{figure6}, the lateral motion of the field line would compress the outer plasma, but the upward motion would contract the field line and reduce the lateral expansion. As a result, the plasma outside the field line is weakly compressed, and the EUV intensity would be weak. Interestingly, although the slow-component EUV wave is weak in the north and south directions, there is still significant difference between the two directions. The slow-component wave is slight stronger in the north direction than in the south direction. We checked the magnetic field inclination angles, and found that, although the field lines in the two directions both are forward-inclined, the inclination angle of the field lines in the south direction is smaller, i.e., the field lines in the south direction are more inclined. Such a subtle difference further validates our paradigm presented in Figure \ref{figure6}. If the field line is extremely forward-inclined, it is expected that no slow-component EUV wave would appear.

In order to confirm our explanation, we performed data-inspired MHD modeling, where several magnetic sources with opposite polarities are imposed below the simulation domain in order to generate magnetic field lines more forward-inclined on one side of the source active region and more backward-inclined on the other side. The simulation results display several interesting features comparable with observations. First, two EUV waves, i.e., a faster wave component and a slower nonwave component, can be clearly identified, which is similar to our previous 3D MHD simulations \cite{Guo2023}. Second, in the data-inspired simulations, the magnetic field lines to the north of the source active region are more backward-inclined, and the associated slow-component EUV waves are more intense. In contrast, the magnetic field lines to the south of the source active region are more forward-inclined, and the corresponding slow-component EUV waves are much weaker. This means that our data-inspired MHD modeling successfully reproduced the correlation between the intensity of the slow-component EUV waves and the exterior angle of the local magnetic field lines.

To summarize, observations of the 8 March 2019 flare/CME event indicate that the event was associated with two types of EUV waves, one is faster and the other is slower. While the fast-component EUV wave is marginally isotropic, the slow-component EUV wave is strongly asnisotropic, being extremely bright in the southwest direction, moderate in the north direction, but much weaker in other directions. We revealed a strong correlation between the brightness of the slow-component EUV wave and the magnetic field configuration. To account for such a correlation, we proposed a model in the framework of the magnetic field-line stretching model: The visibility of the slow-component EUV wave depends on the exterior angle of the magnetic field lines that overlie the source eruption region. The slow-component EUV wave would be brighter if the leg of the magnetic field line is backward-inclined, and it would be weaker if the leg of the magnetic field line is forward-inclined. The more forward-inclined the field line is, the weaker the slow-component EUV wave would be. The data-inspired MHD simulations further validates such a paradigm.

Our results provide a scenario about how to apply EUV waves to diagnose the coronal magnetic fields. For the following slower nonwave component EUV wave, its intensity is highly related to the field-line configuration, i.e., it is more visible at the places with backward-inclined magnetic field lines. With the automated detection of EUV wave features \citep{ireland19, xu20}, it is worthwhile to quantify the relationship between the intensity of EUV waves and their magnetic fields, so as to establish the basis for the EUV waves seismology.

\begin{acknowledgements}
This research was supported by the National Key Research and Development Program of China (2020YFC2201200), NSFC (12127901) and the fellowship of China National Postdoctoral Program for Innovative Talents under grant No. BX20240159. We acknowledge the use of data from SDO. The numerical calculations in this paper were performed in the cluster system of the High Performance Computing Center (HPCC) of Nanjing University. 
\end{acknowledgements}

\bibliographystyle{aa}
\bibliography{ms}

\end{document}